\documentclass[a4paper,11pt]{article}
\usepackage[utf8]{inputenc}
\usepackage[english]{babel}
\usepackage{amsmath, amssymb,graphicx}
\usepackage{braket}
\usepackage{yfonts}
\usepackage{caption}
\usepackage{subcaption}
\usepackage{ulem}
\pdfoutput=1
\usepackage{jheppub}

\title{Proof of the polar decomposition of the Wiener measure}

\author[a]{V.V. Belokurov,}
\author[a]{E.T. Shavgulidze,}
\author[a]{and N.E. Shavgulidze}

\affiliation[a]{Lomonosov Moscow State University,}

\emailAdd{vvbelokurov@yandex.ru}
\emailAdd{shavgulidze@bk.ru}
\emailAdd{nathalia.shavgulidze@math.msu.ru}

\abstract
{In the paper, we give the proof of the polar decomposition of the Wiener measure according to the orbits of the group of diffeomorphisms.\footnote{The research has been performed in the framework of the state assignment of M.V. Lomonosov Moscow State University. }}


\arxivnumber{.}

\begin{document}
\maketitle

\section{ Introduction: Notations and definitions. Formulation of Main Theorem. }

\vspace{0.5cm}

\label{s1}

The paper is devoted to a property of the Wiener measure connected with the action of the group of diffeomorphisms on the space of continuous paths.

Note that the Wiener measure (\cite{(Wiener1)}, \cite{(Wiener2)}, see also, recent books \cite{(Shiryaev)}) is one of the basic concepts of the modern theoretical and mathematical physics.

It is sufficient to note that the Wiener measure has  the important role in quantum physics
 (\cite{(Feynman)}--\cite{(SmSh)}).

The Wiener measure is quasi-invariant with respect to the group of diffeomorphysms (\cite{(Shepp)}) and is used to construct the measures on the groups of diffeomorphisms of the interval and of the circle that are quasi-invariant with respect to the action of smooth subgroups
(\cite{(Shavgulidze1978)} -- \cite{(BSh1)}).

The property of quasi-invariance appears to be very useful for path integrals calculations, and, in particular, for Schwarzian path integrals calculus (see \cite{(BShExact)} -- \cite{(BShCalc)}).

The quasi-invariance of the Wiener measure was used to derive the polar decomposition of the measure and to relate the Schwarzian theory with
conformal quantum mechanics and other theories
( \cite{(BShUnusual)} -- \cite{(ShShPolar)}). 

In the present paper, we give the detailed proof of the polar decomposition of the Wiener measure in the one-dimensional space and demonstrate the generalization of the result to the two-dimensional case.

\textbf{Notations and definitions.}

Let $W^{0}_{\sigma}$ be the Wiener measure with the dispersion $\sigma>0$ on the space
 $$
 C_0([0, 1])=\{x \in C([0, 1]):\ \ x (0)=0 \}\,.
$$

Define the homeomorphism between the spaces $C([0, 1]))$ and $\mathbf{R}\times  C_0([0, 1])$ by the correspondence:
$\forall x\in C([0, 1])$ its image is $\left( x(0), \xi \right)\in \mathbf{R}\times  C_0([0, 1])$ where $\xi(t)=x(t)-x(0)\,, \forall t\in [0, 1]\,.$

Now the Wiener measure $W_{\sigma}$ on the space $C[0,\,1]$ with the help of the above homeomorphism can be defined as the image of the tensorial
product of the Lebeg measure on the real axis $\mathbf{R}$ and the Wiener measure $W^{0}_{\sigma}$ on the space $
 C_0([0, 1])\,.$

 Consider the topological space
  $$
 C_{+}([0, 1])=\{x \in C([0, 1]):\ \ x (t)>0 \}
$$
as the open subset of the space $C([0,\,1])$ with the hereditary topology. Denote the reduction of the Wiener measure $W_{\sigma}$ on the space
$ C_{+}([0, 1])$ as $w_{\sigma}\,.$

 Define the action of the group of diffeomorphisms of the unit interval  $[0,\,1]$
 $$
 Diff^{1}_{+}([0,1])=\{\varphi \in C^{1}([0,1]):\ \  \varphi^{'}(t)>0 \,,\ \
\varphi (0)=0,\, \varphi (1)=1 ,\,\}
$$
on the space $C([0, 1])$ as
\begin{equation}
   \label{graction}
\varphi\, x(t)=x(\varphi ^{-1}(t)) \frac{1}{\sqrt{(\varphi ^{-1}(t))^{'}}}\,,\ \ \ \ \ \ \  \varphi \in Diff^{1}_{+}([0,1])\,, \ \  x \in C([0, 1])\,.
\end{equation}
The group operation being the composition of two diffeomorphisms.

Consider the homeomorphism $A:  Diff^{1}_{+}[0,\,1]) \to C_{0} ([0,\,1])$, of the form
\begin{equation}
   \label{xfi}
   x = A(\varphi)\,:\ \ \ \ \
x(t)=\ln (\varphi'(t))-\ln (\varphi'(0))
\end{equation}
\begin{equation}
   \label{fix}
   \varphi = A^{-1} (x)\,:\ \ \ \ \ \varphi(t)=\frac{\int \limits _{0}^{t}\,e^{x(\tau)}d\tau}
{\int \limits _{0}^{1}\,e^{x(\tau)}d\tau }\,,
\end{equation}
where $\ \ \varphi \in Diff^{1}_{+}([0,1]), \ \ \ \ \  x \in C_0 ([0,1])\,\ \ \ \ \  t \in [0,1]\,.$

The measure $\mu_{\sigma}$ on the group $ Diff^{1}_{+}([0,\,1])$ is the image of the Wiener measure
$W^{0}_{\sigma}(dx)$ on the space $C_{0}([0,\, 1])\,.$ That is, for any Borel subset
$\Psi \subset Diff^{1}_{+}([0,\,1])\,,$ we have $\mu_{\sigma}(\Psi)=W^{0}_{\sigma}(A(\Psi))\,.$

Thus the equality
\begin{equation}
   \label{muW}
\int \limits _{Diff_{+}^{1}([0,1])}
f(\varphi)\,\mu _{\sigma}(d\varphi ) =
\int \limits _{C_{0} ([0,1])}f(A^{-1} (x))\,W^{0}_{\sigma}(d x)
\end{equation}
is valid.

Under the action of the group of diffeomorphisms (\ref{graction}), the space $C_{+}([0, 1])$ is stratified into the orbits that are parametrized
by the positive number $\rho\,:$
$$
\frac{1}{\rho^{2}}=\int \limits_{0}^{1} \frac{dt}{(x(t))^2}.
$$
Note that it is invariant under the group transformations (\ref{graction}) :
$$
\int \limits_{0}^{1} \frac{dt}{(\varphi x(t))^2}=
\int \limits_{0}^{1} \frac{(\varphi^{-1}(t))'dt}{(x(\varphi ^{-1} (t)))^2}=
\int \limits_{0}^{1} \frac{d\tau}{(x(\tau))^2},
$$
where
$\tau =\varphi ^{-1} (t)$.

 Now, define the homeomorphism $B: \mathbf{R}^{+} \times Diff^{1}_{+}([0,1]) \to C_{+}([0,1])\ \ \ (x_{\rho,\varphi}=B(\rho,\varphi ))$ by the equation
\begin{equation}
   \label{xrofi}
x_{\rho,\varphi}(t)=\frac{\rho}{\sqrt{(\varphi ^{-1}(t))'}}\,,
\end{equation}
where $\rho \in \mathbf{R}^{+}=(0,+\infty)$ and
 $\varphi \in Diff^{1}_{+}([0,1]),\,\, t \in [0,1]$.

Then $\forall x \in C_+([0,1])$ we get
$(\rho,\varphi )=B^{-1} (x)\ :$
\begin{equation}
   \label{rofix}
\rho=\left(\int \limits_{0}^{1} \frac{dt}{(x(t))^2} \right)^{-\frac{1}{2}}\,,\ \ \ \
\varphi ^{-1}(t)=\rho^{2} \int \limits_{0}^{t} \frac{d\tau}{(x(\tau))^2}\,.
\end{equation}

Now, we are ready to formulate Main Theorem on the polar decomposition of the Wiener measure.

\textbf{Main Theorem (Theorem 1).}
The Wiener measure is represented as the product of the two measures:
\begin{equation}
   \label{poldecomp1}
w_\sigma(dx)=\exp\{ -\frac{\sigma ^2}{4\rho^2}\}\ (\varphi '(0) \varphi '(1)) ^{\frac{3}{4}} \ \mu_{\frac{2\sigma}{\rho}}(d\varphi)\ d\rho,
\end{equation}
where $\ x\in C_{+}([0,\,1])\,, \ \   \varphi\in Diff^{1}_{+}([0,\,1])\ $ and $\ \  0<\rho<+\infty\ $ are related by eqs. (\ref{xrofi}), (\ref{rofix}).

\section{Proof of Main Theorem.}

\vspace{0.5cm}

To prove the \textbf{Theorem 1}, we formulate and prove several auxiliary theorems and lemmas.

\textbf{Theorem 2.}
For $x\,,\ \ \varphi\,, \ \ \rho\  $ from the theorem 1 and for
$\theta >0\,,\ $ we have
$$
\int \limits _{C_{+} ([0,1])}\ \delta\left(x(1)-\theta x(0)\right)\,
e^{-\frac{a^2}{\sigma ^2} x^2(0)}\ w_\sigma (d x)=
$$
$$=\int \limits _{0}^{+\infty}\ \int \limits _{Diff_{+}^{1}([0,1])}\,
\delta\left(\rho\sqrt{\varphi'(1)}-\theta \rho\sqrt{\varphi'(0)}\right)\,
\exp\left\{-\frac{a^2}{\sigma ^2}\rho^2\varphi'(0)-\frac{1}{8\rho^2}\right\}
(\varphi'(0)\varphi'(1))^{\frac{3}{4}}\ \mu _{\frac{2\sigma}{\rho}}(d\varphi )\ d\rho\,.
$$

\textbf{Lemma 1.}
$$\int \limits _{C_+ ([0,1])}\ \delta\left(x(1)-x(0)\right)\,e^{-\frac{a^2}{\sigma ^2} x^2(0)} w_\sigma (dx)=
\frac{1}{2\sqrt{2 }}	\left(	\frac{1}{a}	-	\frac{1}{\sqrt{a^2+2}}	\right).$$

\textbf{Proof of Lemma 1.}
As
$$\int \limits _{C([0,\,1])}F(x(0), x(1))W_\sigma(d x) =\frac{1}{\sqrt{2\pi }\sigma}
\int \limits _{-\infty}^{+\infty}	\int \limits _{-\infty}^{+\infty}F(q_0,q_1)
e^{-\frac{(q_1-q_0)^{2}}{2 \sigma ^{2}}}\ dq_0\, dq_1$$
and
$$\int \limits _{x(0)\geq 0, x(1)\geq 0, x\notin C_+ ([0,\,1])}F(x(0), x(1))W_\sigma(d x)
=\frac{1}{\sqrt{2\pi }\sigma}
\int \limits _{0}^{+\infty}	\int \limits _{0}^{+\infty}F(q_0,q_1)
e^{-\frac{(q_1+q_0)^2}{2 \sigma ^2}}dq_0 \,dq_1\,,$$
then
$$\int \limits _{C_+ ([0,1])}\ \delta\left(x(1)-x(0)\right)\,
e^{-\frac{a^2}{\sigma ^2} x^2(0)} w_\sigma(dx)=$$
$$=\frac{1}{\sqrt{2\pi }\sigma} \int \limits _{0}^{+\infty}	\int \limits _{0}^{+\infty}\delta(q_1-q_0)
e^{-\frac{a^2}{\sigma ^2} q_0^2}
(e^{-\frac{(q_1-q_0)^2}{2 \sigma ^2}}-e^{-\frac{(q_1+q_0)^2}{2 \sigma ^2}})dq_0 dq_1=$$
$$=\frac{1}{\sqrt{2\pi \sigma}} \int \limits _{0}^{+\infty}	\int \limits _{0}^{+\infty}
\delta(q_1-q_0)(e^{-\frac{a^2}{\sigma ^2} q_0^2}-e^{-\frac{a^2+2}{\sigma ^2} q_0^2})dq_0\, dq_1
=\frac{1}{\sqrt{2\pi \sigma}} \int \limits _{0}^{+\infty}
(e^{-\frac{a^2}{\sigma ^2} q_0^2}-e^{-\frac{a^2+2}{\sigma ^2} q_0^2})dq_0=$$
$$=\frac{1}{\sqrt{2\pi }}	\left(	\frac{\sqrt{\pi}}{2a}	-	\frac{\sqrt{\pi}}{2\sqrt{a^2+2}}	\right)
=\frac{1}{2\sqrt{2 }}	\left(	\frac{1}{a}	-	\frac{1}{\sqrt{a^2+2}}	\right).$$

Consider the subgroup
$$
Diff^{3}_{+}([0,1])=Diff^{1}_{+}([0,1])\cap C^{3}([0,1]).
$$
For any Borel subset $\Psi\subset Diff^{1}_{+}([0,\,1])$ and $g \in Diff^{3}_{+}([0,\,1])\,,$
define $g \Psi = \{ g \circ \varphi: \varphi \in \Psi \}$ and
$\mu^{g} _{\sigma}(\Psi)=\mu_{\sigma}(g \Psi)\,.$

The following theorem is valid

\textbf{Theorem 3.}
For any Borel subset $\Psi\subset Diff^{1}_{+}([0,\,1])$ and $g \in Diff^{3}_{+}([0,\,1])\,,$
\begin{equation}
   \label{QuasInv}
\mu ^{g} _{\sigma}(\Psi) =
\int \limits _{\Psi}\ p_{g}(\varphi)\,\mu_{\sigma}(d\varphi).
\end{equation}
Here,
$$
p_{g}(\varphi)=\frac{d \mu_{\sigma}^{g}}{d \mu_{\sigma}} (\varphi)=
$$
\begin{equation}
   \label{RadNik}
=\frac{1}{\sqrt{g'(0)g'(1)}}\exp\left\{	\frac1{\sigma^2} \left(	\frac{g''(0)}{g'(0)}\varphi'(0) - \frac{g''(1)}{g'(1)}\varphi'(1)	\right)
+ \frac{1}{\sigma^2} \int \limits _{0}^{1} Sch\{g,\,\varphi(\tau)\} (\varphi'(\tau))^2 d\tau	\right\}
\end{equation}
is the Radon-Nikodim derivative, and
$ Sch\{g,\,t\}=\left(\frac{g''(t)}{g'(t)}\right)'-\frac{1}{2}\left(	\frac{g''(t)}{g'(t)}\right)^2 $  is the Schwarzian derivative.

Note that $\ \mu_{\sigma}(d\varphi)\ $ is the quasi-invariant measure on $\ Diff^{1}_{+}([0,\,1])\ $ under the action of the subgroup
$\ Diff^{3}_{+}([0,\,1])\,,\ $ that is
$\
\mu^{g}_{\sigma}(d\varphi)=\mu_{\sigma}(d(g\circ \varphi))=p_g(\varphi)\mu_{\sigma}(d\varphi)\,,\ \
$
for $g \in Diff^{3}_{+}([0,\,1])\,.$

 \textbf{ Lemma 2.}
$p_{g_{\beta}}(\varphi)=
\exp\{\frac{2\beta}{\sigma^2} (-\varphi'(0)+\frac1{\beta+1}\varphi'(1))\}$

For any Borel subset $X\subset Diff^{1}_{+}([0,\,1])\,,\ $ and the function $g_{\beta}(t)=\frac{(\beta+1)t}{\beta t +1}$ where $\beta\in \mathbf{R}\,, \ \ \beta>-1\,, (\ g_{\beta} \in Diff^{3}_{+}([0,\,1])\,)$
the equality
\begin{equation}
   \label{lemma2}
\mu ^{g_{\beta}}_{\sigma}(X) =
\int \limits _{X}\ \exp\{\frac{2\beta}{\sigma^2} (-\varphi'(0)+\frac1{\beta+1}\varphi'(1))\}\
\mu_{\sigma}(d\varphi)
\end{equation}
is valid.

\textbf{Proof of Lemma 2.}
For the function $g_{\beta}$, we have
$$g_{\beta}'(t)=\frac{\beta+1}{(\beta t +1)^2}\,,\ \ \
g_{\beta}''(t)=-\frac{2\beta(\beta+1)}{(\beta t +1)^3}\,,\ \ \
\frac{g_{\beta}''(t)}{g_{\beta}'(t)}=-\frac{2\beta}{\beta t +1}
$$
and
$$
\mathcal{Sch}\{g_{\beta}\,,\ t\}= \left(\frac{g_{\beta}''(t)}{g_{\beta}'(t)}\right)'
-\frac{1}{2} \left(\frac{g_{\beta}''(t)}{g_{\beta}'(t)}\right)^2=
\frac{2\beta ^2}{(\beta t +1)^2}-\frac{1}{2}\left(-\frac{2\beta}{\beta t +1}\right)^2=0\,.
$$

From (\ref{RadNik}), it follows that
$$
p_{g_{\beta}}(\varphi)=\exp\{\frac{2\beta}{\sigma^2} (-\varphi'(0)+\frac1{\beta+1}\varphi'(1))\}\,,
$$
that proves the Lemma 2.

Consider the integral
\begin{equation}
   \label{jalfa}
J(\alpha)=\int \limits _{Diff_{+}^{1}([0,1])}\
\delta\left(\sqrt{\frac{\varphi'(1)}{\varphi'(0)}}-1\right)\exp\left\{-\frac{\alpha}{\sigma^2} \rho^2\varphi'(0)\right\}\left(\frac{\varphi'(0)}{\varphi'(1)}\right)^{\frac{3}{4}}\,
\mu _{\frac{2\sigma }{\rho}}(d\varphi )=
\end{equation}
$$=\int \limits _{Diff_{+}^{1}([0,1])}\
\delta\left(\sqrt{\frac{\varphi'(1)}{\varphi'(0)}}-1\right)\exp\left\{-\frac{\alpha}{\sigma^2} \rho^2\varphi'(0)\right\}\,\mu _{\frac{2\sigma }{\rho}}(d\varphi ).$$

\textbf{ Lemma 3.}
For $\alpha=\frac{\beta^2}{2(\beta+1)}\,,$ and $\beta>-1\,,$
\begin{equation}
   \label{lemma3}
J\left(\frac{\beta^2}{2(\beta+1)}\right)
=\frac{\rho}{\sqrt{2\pi}\sigma}	\exp\left\{ -\frac{\rho^2}{2\sigma ^2} \ln^2(\beta+1) \right\}\,.
\end{equation}

\textbf{Proof of Lemma 3.}
Define
$$
\psi=g_{\beta}\circ\varphi\,,\ \ \ \ \
\psi(\tau)=\frac{(\beta+1)\varphi(\tau)}{\beta \varphi(\tau) +1}\,.
$$

Since
$$\psi'(\tau)=\frac{(\beta+1)\varphi'(\tau)}{(\beta \varphi(\tau) +1)^2}\,, \ \psi'(0)=(\beta+1)\varphi'(0)\,, \ \psi'(1)=\frac1{\beta+1}\varphi'(1)\,,
$$
then
$$
\delta\left(	\sqrt{\frac{\psi'(1)}{\psi'(0)}} -\frac1{\beta+1}	\right)=\delta\left(	\frac1{\beta+1}\sqrt{\frac{\varphi'(1)}{\varphi'(0)}} -\frac1{\beta+1}	\right)=
(\beta+1)\,\delta\left(	\sqrt{\frac{\varphi'(1)}{\varphi'(0)}} -1	\right)
$$
and
$$
\int \limits _{Diff_+^1([0,1])}\	\delta\left(	\sqrt{\frac{\psi'(1)}{\psi'(0)}} -\frac1{\beta+1}	\right)	\mu_{\frac{2\sigma}{\rho}}(d\psi) =
$$
$$
=\int \limits _{Diff_+^1([0,1])}\		(\beta+1)\,\delta\left(	\sqrt{\frac{\varphi'(1)}{\varphi'(0)}} -1	\right)	\exp\left\{\frac{ \rho^2 \beta}{2\sigma^2} \left(-\varphi'(0)+\frac1{\beta+1}\varphi'(1)\right)\right\}\mu_{\frac{2\sigma}{\rho}}(d\varphi)=
$$
$$
=(\beta+1)\int \limits _{Diff_+^1([0,1])}\!\!		\delta\left(	\sqrt{\frac{\varphi'(1)}{\varphi'(0)}} -1	\right)	\exp\left\{-\frac{\rho^2\beta^2\varphi'(0)}{2\sigma ^2 (\beta+1)}
\right\}\mu_{\frac{2\sigma}{\rho}}(d\varphi)=
$$	
$$
=(\beta+1)\ J\left(\frac{\beta^2}{2(\beta+1)}\right).
$$

Thus,
$$
J\left(\frac{\beta^2}{2(\beta+1)}\right)=\frac1{\beta+1} \int \limits _{Diff_+^1([0,1])}\!\!	\delta\left(	\sqrt{\frac{\psi'(1)}{\psi'(0)}} -\frac1{\beta+1}	\right)	
\mu_{\frac{2\sigma}{\rho}}(d\psi)\,.
$$

To calculate the integral, we use the equality
\begin{equation}
   \label{26}
F : Diff^{1}_{+}([0,\,1])\to \mathbf{R}\,,\ \ \ \ \ \int \limits _{Diff_+^1([0,1])}
\ F(\psi)\mu _{\frac{2\sigma}{\rho}}(d\psi )\ =\ \int \limits _{C_0([0,\,1])}
\ F(A^{-1}(\xi))W^0 _{\frac{2\sigma}{\rho}}(d\xi )\,.
\end{equation}
Here,
$$
A :	Diff^{1}_{+}([0,\,1])	\to	C_0([0,\,1])\,,\ \ \ \ \
\xi=A(\psi),\ \ \	\xi(t)=\ln\frac{\psi'(t)}{\psi'(0)},\ \ \ \ \frac{\psi'(1)}{\psi'(0)}=e^{\xi(1)}\,.
$$

In particular,
$$
\int \limits _{Diff_+^1([0,1])}\	\delta\left(	\sqrt{\frac{\psi'(1)}{\psi'(0)}} -\frac1{\beta+1}	\right)	\mu_{\frac{2\sigma}{\rho}}(d\psi)=
\ \int \limits _{C_0([0,\,1])}\	\delta\left(	e^{\frac{\xi(1)}2} -\frac1{\beta+1}	\right)
W^0 _{\frac{2\sigma}{\rho}}(d\xi )\,.
$$

Using the properties of $\delta-$ function and the equality
\begin{equation}
   \label{27}
\int \limits _{C_0([0,\,1])}
\ f(\xi(1)\,W^0 _{\sigma}(d\xi )\,=
\frac{1}{\sqrt{2\pi }\sigma}
\int \limits _{-\infty}^{+\infty}	f(q)\,
e^{-\frac{(q)^{2}}{2 \sigma ^{2}}}\ dq\,,
\end{equation}
we obtain
$$\int \limits _{C_0([0,\,1])}\	\delta\left(	e^{\frac{\xi(1)}2} -\frac{1}{\beta+1}	\right)W^0 _{\frac{2\sigma}{\rho}}(d\xi )=
\frac {\rho}{2\sqrt{2\pi}\sigma}\int \limits _{-\infty}^{+\infty}\delta\left(	e^{\frac{q}2} -\frac1{\beta+1}	\right) e^{-\frac{\rho^2 q^2}{2\sigma^2}}dq=$$
$$=\frac{\rho}{2\sqrt{2\pi}\sigma}\int \limits _{-\infty}^{+\infty}
\delta\left(	e^{\frac{q}2} -\frac1{\beta+1}	\right) e^{-\frac{\rho^2 q^2}{8\sigma ^2}}dq
=\frac{\rho}{2\sqrt{2\pi}\sigma}\int \limits _{0}^{+\infty}\delta\left(	y -\frac1{\beta+1}	\right)
e^{-\frac{\rho^2\ln^2y}{2\sigma ^2}}  \, \, \frac2{y}dy=$$
$$=\frac{\rho(\beta+1)}{\sqrt{2\pi}\sigma}\exp\left\{-\frac{\rho^2\ln^2(\beta+1)}{2\sigma^2}\right\},$$
where $y=e^{\frac{q}2}$.
Therefore, (\ref{lemma3}) is proven.

\textbf{ Lemma 4.}
 Let $a^2=\frac{\beta^2}{2(\beta+1)}$, $\beta>-1$.

 Then the following equality is valid
$$
\int \limits _{Diff_+^1([0,1])}\
\delta\left(\rho\sqrt{\varphi'(1)}-\rho\sqrt{\varphi'(0)}\right)
\exp\left\{-\frac{a^2}{\sigma ^2}\rho^{2}\varphi'(0)\right\}(\varphi'(0)\varphi'(1))^{\frac34}
\mu _{\frac{2\sigma}{r}}(d\varphi )=
$$
\begin{equation}
   \label{lemma4}
   =\sqrt{\frac2{\pi}}\frac{(\beta+1)\ln(\beta+1)}{\sigma\beta(\beta+2)}
\exp\left\{ -\frac{\rho^2}{2\sigma ^2} \ln^2(\beta+1) \right\}\,.
\end{equation}

\textbf{Proof of Lemma 4.}
Denote
  $$
  I_1=\int \limits _{Diff_+^1([0,1])}\
\delta\left(\rho\sqrt{\varphi'(1)}-\rho\sqrt{\varphi'(0)}\right)
\exp\left\{-\frac{\alpha}{\sigma ^2} \rho^2\varphi'(0)\right\}(\varphi'(0)\varphi'(1))^{\frac34}\mu _{\frac{2\sigma}{\rho}}(d\varphi )\,.
$$

Due to
$$
\delta\left(\rho\sqrt{\varphi'(1)}-\rho\sqrt{\varphi'(0)}\right)=
\frac{1}{\rho\varphi'(0)}\delta\left(\sqrt{\frac{\varphi'(1)}{\varphi'(0)}}-1\right)\,,$$
the integral
$$
I_1=\ \frac1{\rho}\ \int \limits _{Diff_+^1([0,1])}\
\delta\left(\sqrt{\frac{\varphi'(1)}{\varphi'(0)}}-1\right)
\exp\left\{-\frac{\alpha}{\sigma ^2}\rho^2\varphi'(0)\right\}(\varphi'(0))^{\frac14}(\varphi'(1))^{\frac34}\mu _{\frac{2\sigma}{\rho}}(d\varphi )\,.
$$

From (\ref{jalfa}), it follows that
$$
J'(\alpha)=\int \limits _{Diff_+^1([0,1])}\
\delta\left(\sqrt{\frac{\varphi'(1)}{\varphi'(0)}}-1\right)
(-\frac{\rho^2}{\sigma ^2}\varphi'(0))\, \exp{-\frac{\alpha}{\sigma ^2} \rho^2\,\varphi'(0)}
\left(\frac{\varphi'(0)}{\varphi'(1)}\right)^{\frac{3}{4}}\mu _{\frac{2\sigma}{\rho}}(d\varphi )=
	-\frac{\rho^3}{\sigma ^2}\, I_{1}\,.
$$
From the other hand,
$$
J'\left(\frac{\beta^2}{2(\beta+1)}\right)=
\frac{\left(J\left(\frac{\beta^2}{2(\beta+1)}\right)\right)_{\beta}'}	{\left(\frac{\beta^2}{2(\beta+1)}\right)_{\beta}'}=-\sqrt{\frac2{\pi}}\frac{\rho^3(\beta+1)\ln(\beta+1)}{\sigma ^3\beta(\beta+2)}\exp\left\{ -\frac{\rho^2}{2\sigma ^2} \ln^2(\beta+1) \right\}\,.
$$
Now, after the substitution  $\alpha=\frac{\beta^2}{2(\beta+1)}$ in $I_1\,,$ we get  (\ref{lemma4}) that proves  Lemma 4.

Consider the integral
\begin{equation}
   \label{I2}
I_{2}=\int \limits _{0}^{+\infty}\ \int \limits _{Diff_+^1([0,1])}\!\!
\delta\left(\rho\sqrt{\varphi'(1)}-\rho\sqrt{\varphi'(0)}\right)\exp\left\{-a^{2}\, \rho^{2}\,\varphi'(0)-
\frac{\sigma ^{2}}{8\rho^{2}}\right\}(\varphi'(0)\varphi'(1))^{\frac{3}{4}}\,\mu _{\frac{2\sigma}{\rho}}(d\varphi )\, d\rho\,.
\end{equation}
To calculate  (\ref{I2}), we use the results for the following auxiliary integrals:
\begin{equation}
   \label{Ia}
 I_a=\int \limits _{0}^{+\infty}\left(\frac1{\varrho^2}+a\right)\exp\left\{-\frac{b^2}2\left(\frac1{\varrho^2}+a^2\varrho^2\right)\right\}d\varrho=
\frac{\sqrt{2\pi}}{b}e^{-ab^2}\,,
\end{equation}
\begin{equation}
   \label{Ib}
    I_b=\int \limits _{0}^{+\infty}\frac1{\varrho^2}\exp\left\{-\frac{b^2}2\left(\frac1{\varrho^2}+a^2\varrho^2\right)\right\}d\varrho=
\frac{\sqrt{2\pi}}{2b}e^{-ab^2}\,,
\end{equation}
\begin{equation}
   \label{Ic}
 I_c=\int \limits _{0}^{+\infty}\exp\left\{-\frac{b^2}2\left(\frac1{\varrho^2}+a^2\varrho^2\right)\right\}d\varrho=
\frac{\sqrt{2\pi}}{2ab}e^{-ab^2}\,,
\end{equation}
that can be easily obtained with the substitutions $t=a\varrho-\frac{1}{\varrho}$ for $I_{a}\,,$   and $\varrho=\frac{1}{a\rho}$ for $I_{b}\,.$

Using (\ref{lemma4}), (\ref{Ia}), (\ref{Ib}), and (\ref{Ic}),
 we obtain
$$
I_2=\int \limits _{0}^{+\infty}\sqrt{\frac2{\pi}}\frac{(\beta+1)\ln(\beta+1)}{\sigma\beta(\beta+2)}\exp\left\{ -\frac{\rho^2}{2\sigma ^2} \ln^2(\beta+1) \right\}e^{-\frac{\sigma ^2}{8\rho^2}}\,d\rho=
$$
$$
=\int \limits _{0}^{+\infty}\sqrt{\frac2{\pi}}\frac{(\beta+1)\ln(\beta+1)}{\beta(\beta+2)}\exp\left\{ -\frac{\varrho^2}{2} \ln^2(\beta+1) \right\}e^{-\frac1{8\varrho^2}}\,d\varrho=
$$
$$
=\sqrt{\frac2{\pi}}\frac{(\beta+1)\ln(\beta+1)}{\beta(\beta+2)}\int \limits _{0}^{+\infty}	\exp\left\{ -\frac1{8\varrho^2}-\frac{\varrho^2}{2} \ln^2(\beta+1) \right\}d\varrho=
$$
\begin{equation}
   \label{I2result}
=\sqrt{\frac2{\pi}}\frac{(\beta+1)\ln(\beta+1)}{\beta(\beta+2)}		\frac{\sqrt{2\pi}}{2\ln(\beta+1)}e^{-\frac12\ln(\beta+1)}=
\frac{\sqrt{\beta+1}}{\beta(\beta+2)}\,.
\end{equation}

\textbf{Proof of Theorem 2.}
Due to Lemma 1  with
\begin{equation}
   \label{a2}
a^2=\frac{\beta^2}{2(\beta+1)}\,, \ \ \beta>-1\,,
\end{equation}
for $x\in C_{+}([0,\,1])
\,, \ \ \  x(t)=\frac{\rho}{\sqrt{\varphi^{-1}(t))'}}\,,
\ \ \  \varphi\in Diff^{1}_{+}([0,\,1])\,,\ \ \ \  0<\rho<+\infty\,,$ we have
$$
\int \limits _{C_+ ([0,1])}\, \delta\left(x(1)-x(0)\right)\,
e^{-\frac{a^2}{\sigma ^2} x^2(0)} w_\sigma (d x)=\frac{1}{2\sqrt{2}}\left(\frac{1}{a}-\frac{1}{\sqrt{a^{2}+2}} \right)=I_{2}\,.
$$
Thus,
$$
I_{2}=\int \limits _{0}^{+\infty}\ \int \limits _{Diff_{+}^{1}([0,1])}\,
\delta\left(\rho\sqrt{\varphi'(1)}-\rho\sqrt{\varphi'(0)}\right)
\exp\left\{-\frac{a^2}{\sigma ^2}\rho^2\varphi'(0)-\frac{\sigma ^{2}}{8\rho^{2}}\right\}
(\varphi'(0)\varphi'(1))^{\frac{3}{4}}\,\mu _{\frac{2\sigma}{\rho}}(d\varphi )\, d\rho\,.
$$

Let   $y(t)$  and $\psi (t)$ be
$$
y(t)=g_{\beta} x(t)=x(g_{\beta} ^{-1}(t)) \frac{1}{\sqrt{(g_{\beta} ^{-1}(t))'}}\,,\ \ \ \
\psi (t)=g_{\beta} (\varphi (t))\,.
$$
In this case, we get (see \cite{(Shepp)})
$$
\int \limits _{C_+ ([0,1])}\ \delta\left(\frac{1}{\beta+1}x(1)-(\beta+1) x(0)\right)\,
\exp\{-\frac{\alpha}{\sigma ^2} x^2(0)+\frac{2\beta}{\sigma^2} (-x^2 (0)+\frac1{\beta+1}x^2(1))\}\, w_\sigma (d x)=
$$
$$
=\int \limits _{C_+ ([0,1])}\ \,\delta\left(y(1)- y(0)\right)\,
e^{-\frac{\alpha}{\sigma ^2} y^2(0)} w_\sigma (d y)=
$$
$$
=\int \limits _{0}^{+\infty}\ \int \limits _{Diff_+^1([0,1])}\
\delta\left(\rho\sqrt{\psi'(1)}- \rho\sqrt{\psi'(0)}\right)
\exp\left\{-\frac{\alpha}{\sigma ^2}\rho^2\psi'(0)-\frac{1}{8\rho^2}\right\}
(\psi'(0)\psi'(1))^{\frac{3}{4}}\,\mu _{\frac{2\sigma}{\rho}}(d\psi )\, d\rho=
$$
$$
=\int \limits _{0}^{+\infty}\ \int \limits _{Diff_+^1([0,1])}\
\delta\left(\frac{1}{\beta+1}\,\rho\,\sqrt{\varphi'(1)}-(\beta+1)\,\rho\,\sqrt{\varphi'(0)}\right)\times
$$
$$
\times\exp\left\{-\frac{\alpha}{\sigma ^2}\rho^2\varphi'(0)+\frac{\rho^2\beta}{2\sigma^2} (-\varphi'(0)+\frac1{\beta+1}\varphi'(1))-\frac{1}{8\rho^2}\right\}
(\varphi'(0)\varphi'(1))^{\frac{3}{4}}\mu _{\frac{2\sigma}{\rho}}(d\varphi )\, d\rho\,,
$$
that proves Theorem 2.

Now we are going to prove Theorem 1. First, we are to prove several lemmas.

For any $u\in C([0,\,1])\,,$ define
$ \left\| u \right\|= \max \limits _{t \in [0,1]} \left| u(t) \right|.$

\textbf{ Lemma 5.}
$$
\forall v\ \in\, C([0,\,1])\,,\ \   \left\| v \right\| \leq \frac{1}{4}\,,\ \ \ \ \
\exists f \ \in\, Diff^{3}_{+}([0,1])\,:
$$
$$
\mathcal{S}ch\, \{f\,,\ t\}=
\left(\frac{f''(t)}{f'(t)}\right)'
-\frac{1}{2}\left(\frac{f''(t)}{f'(t)}\right)^2\,=v(t)\,,
$$
with $f''(1)=0, \ $ and $\ \left|\frac{f''(0)}{f'(0)}\right|\leq \frac{1}{2}\,.$

\textbf{Proof of Lemma 5.} Define the space $X$ as
$$
X=\{u\in C([0,\,1]): \left\| u \right\| \leq  \frac{1}{2}\}
\,,
$$
and consider the map $Q_{v} : X \to X $
\begin{equation}
   \label{Qv}
Q_v (u) (t)= v(t)+ \frac{1}{2} (\int \limits_{1-t}^1 u(\tau) d\tau )^2\,,
u \in X\,.
\end{equation}
Since $(\int \limits_{1-t}^{1} \left|u(t) \right| dt )^2\leq \frac{1}{4}\,,$ then
 $ \left\| Q_v (u) \right\| \leq \frac{1}{2}$ and $Q_v (u) \in X\,.$

 Note that for any $u_1\, ,\, u_2 \in X\,,$ we have
 $$
 Q_v (u_1) (t) - Q_v (u_1) (t)= \frac{1}{2} (\int \limits_{1-t}^1 (u_2(\tau) + u_1(\tau))d\tau )
(\int \limits_{1-t}^1 (u_2(\tau) - u_1(\tau))d\tau  )
$$
and
$$
\left\|Q_v (u_2)  - Q_v (u_1)\right\| \leq
\frac{1}{2} (\int \limits_0^1 (\left\|u_2\right\| + \left\|u_1 \right\|))d\tau )
(\int \limits_0^1 \left\| u_2 - u_1 \right\|d\tau  )\leq
$$
$$
\frac{1}{2}(\left\|u_2\right\| + \left\|u_1 \right\|)
( \left\| u_2 - u_1 \right\| ) \leq
\frac{1}{2}( \left\| u_2 - u_1 \right\| )\,.
$$

Therefore, the map $Q_{v}$ is the contractive map with respect to the metric given by the norm $\|\,.\,\|\,.$
The set $X$ is closed in the space $C\left([0,\,1]\right)\,. $ Hence it is a complete metric space. Due to the theorem on a contractive map, there
exists $u\,\in\,X\,$ that
$u=Q_v(u)\,,$ that is
$$
u (t) -\frac{1}{2} \left(\int \limits_{1-t}^1 u(\tau) d\tau \right)^2= v(t)\,.
$$

Define
$$
f (t) =\frac { \int \limits_0^t \exp \left\{ \int \limits_0^{\tau}
\left(\int \limits_{1-\tau_1}^1 u(\tau_2)\,d\tau_2\,\right)\, d\tau_1\right\}\, d\tau }
{ \int \limits_0^1 \exp \left \{ \int \limits_0^{\tau}
\left(\int \limits_{1-\tau_1}^1 u(\tau_2)\,d\tau_2\,\right)\, d\tau_1 \right\}\, d\tau }\,.$$
In this case,
$$
\frac{f''(t)}{f'(t)}=\int \limits_{1-t}^1 u(\tau) d\tau\,, \ \
f''(1)=0\,, \ \    \left|\frac{f''(0)}{f'(0)}\right|\leq \frac{1}{2}\,, \ \
\left(\frac{f''(t)}{f'(t)}\right)'=u(t)\,,
$$
and
$$
\left(\frac{f''(t)}{f'(t)}\right)'
-\frac{1}{2}\left(\frac{f''(t)}{f'(t)}\right)^2\,=\,v(t)\,.
$$
That proves Lemma5.

For any subspace $Y \subset C ([0,1])\,$ define
$$
X(Y)=\{x \in C_{+} ([0,1]): \ x^2 \in Y\}\,,\ \ \ x^2=(x(t))^2 \,.
$$

For any Borel set $Y \subset C ([0,1])\,,$ consider Borel measures
$$
\mu _1(Y)=\int \limits _{X(Y)}\ \delta\left(x(1)- x(0)\right)\,
\exp \left\{-\frac{a^2}{\sigma ^2} x^2(0)\right\}\, w_\sigma (d x)
$$
and
$$
\mu _2(Y)=\int \limits _{0}^{+\infty}\ \int \limits _{Diff_+^1([0,1])}\
\chi _{X(Y)} \left(x_{\rho, \varphi}\right)\,
\delta\left(\rho\sqrt{\varphi'(1)}- \rho\sqrt{\varphi'(0)}\right)
$$
$$
\exp\left\{-\frac{a^2}{\sigma ^2}\rho^2\varphi'(0)-\frac{\sigma ^2}{8\varrho^2}\right\}
\left(\varphi'(0)\varphi'(1)\right)^{\frac{3}{4}}\ \mu _{\frac{2\sigma}{\rho}}(d\varphi )\, d\rho\,.
$$
Here, $\chi _{X(Y)} (x)$ is the characteristic function of the set $X(Y)$, that is $\chi _{X(Y)} (x)\,=\,1\,,$ if
$x \in X(Y)\,,$  $\chi _{X(Y)} (x)\,=\,0\,,$,  $x \notin X(Y)\,,$ and
$$
x_{\rho, \varphi}\in C_{+}([0,\,1])
\,, \ \ \ \ x_{\rho, \varphi}(t)=\frac{\rho}{\sqrt{\varphi^{-1}(t))'}}\,,
\ \ \ \ \ \varphi\in Diff^{1}_{+}([0,\,1])\,,\ \ \ \  0<\rho<+\infty\,.
$$

\textbf{Lemma 6.} For any $v\in C([0,\,1])\,, \left\| v \right\| \leq \frac{1}{4}\,,$
$$
I_{3}\equiv\int \limits _{C([0,\,1])} \exp\left\{-\frac{1}{4\sigma ^2}\int \limits_{0}^{1} v(t)\,y(t)\,,dt  \right\}\,\mu _1(dy)
=\int \limits _{C([0,\,1])} \exp\left\{-\frac{1}{4\sigma ^2}\int \limits_{0}^{1} v(t)\,y(t)\,,dt  \right\}\,\mu _2(dy)\equiv I_{4}\,.
$$

\textbf{Proof of Lemma 6.} According to Lemma 5, we have
$$
\int \limits _{C([0,\,1])}\ e^{-\frac{1}{4\sigma ^2}\int \limits_{0}^1 v(t)y (t)dt  }\mu _1(dy)=
\int \limits _{C_+ ([0,1])}\ \delta\left(x(1)- x(0)\right)\,
e^{-\frac{1}{4\sigma ^2}\int \limits_{0}^1 v(t)x^2(t)dt  }
e^{-\frac{a^2}{\sigma ^2} x^2(0)} w_\sigma (d x)=
$$
$$
=\frac{1}{\sqrt[4]{f'(0)f'(1)}}
\int \limits _{C_+ ([0,1])}\
\delta\left(\frac{1}{\sqrt{f'(1)}}(\sqrt{f'(1)}x(1))- \frac{1}{\sqrt{f'(0)}}(\sqrt{f'(0)}x(0))\right)\times
$$
$$
\times\exp\left\{-\frac{1}{f'(0)\sigma ^2} (a^2+\frac{f''(0)}{4f'(0)})(f'(0)x^2(0))\right\}\times
$$
$$
\times\exp\left\{	\frac1{4\sigma^2} \left(	\frac{f''(0)}{f'(0)}x^2(0)
 - \frac{f''(1)}{f'(1)}x^2 (1)	\right)
+ \frac{1}{4\sigma^2} \int \limits _{0}^{1}\,\mathcal{S}ch \left\{f\,,\,t\right\}\, (x(t))^{2}\, dt	\right\}
w_\sigma (d x)=
$$
$$
=\frac{1}{\sqrt[4]{f'(0)f'(1)}}
\int \limits _{C_+ ([0,1])}\!\!\!
\delta\left(\frac{1}{\sqrt{f'(1)}}u(1)- \frac{1}{\sqrt{f'(0)}}u(0)\right)\,
\exp\left\{-\frac{1}{f'(0)\sigma ^2} (a^2+\frac{f''(0)}{4f'(0)})u^2(0)\right\}
w_\sigma (du)\,,
$$
where $u(\tau)=\frac{x(f^{-1}(t))}{\sqrt{(f^{-1}(t))'}}$.

From the other side,
$$
\int \limits _{C([0,\,1])}\ \exp\left\{-\frac{1}{4\sigma ^2}\int \limits_{0}^{1} v(t)\,y(t)\,dt \right \}\ \mu _{2}(dy)=
$$
$$
=\int \limits _{0}^{+\infty}\ \int \limits _{Diff_{+}^{1}([0,1])}\
\exp\left\{-\frac{1}{4\sigma ^2}\int \limits_{0}^{1} v(\tau)\,\frac{\rho^2}{(\varphi^{-1}(\tau))'}\,d\tau  \right\}
\delta\left(\rho\sqrt{\varphi'(1)}- \rho\sqrt{\varphi'(0)}\right)\times
$$
$$
\times\exp\left\{-\frac{a^2}{\sigma ^2}\,\rho^2\,\varphi'(0)-\frac{\sigma ^2}{8\rho^2}\right\}
(\varphi'(0)\varphi'(1))^{\frac{3}{4}}\,\mu _{\frac{2\sigma}{\rho}}(d\varphi )\, d\rho\,.
$$
If $\tau=\varphi(t)\,,$ than
$$
\int \limits_{0}^{1} v(\tau)\,\frac{\rho^2}{(\varphi^{-1}(\tau))'}\,d\tau=
\rho^{2}\int \limits_{0}^1 \, \mathcal{S}ch\,\left\{f\,,\, \varphi (t)\right\}\,(\varphi'(t))^{2}\,dt\,,
$$

Therefore,
$$
\int \limits _{C([0,\,1])}\ \exp\left\{-\frac{1}{4\sigma ^2}\int \limits_{0}^{1}\ v(t)\,y(t)\,dt  \right\}\ \mu _{2}(dy)=
$$
$$
=\frac{1}{\sqrt[4]{f'(0)f'(1)}}
\int \limits _{0}^{+\infty}\ \exp\left\{-\frac{\sigma ^2}{8\rho^2}\right\}\times
$$
$$
\times\int \limits _{Diff_+^1([0,1])}\ \delta \left(\frac{1}{\sqrt{f'(1)}}\,\rho\,\sqrt{f'(1)\varphi'(1)}-
\frac{1}{\sqrt{f'(0)}}\,\rho\, \sqrt{f'(0)\varphi'(0)}\right)\times
$$
$$
\times\exp\left\{-\frac{1}{f'(0)\sigma ^2}(a^2+\frac{f''(0)}{4f'(0)} )\,\rho^2\, f'(0)\varphi'(0)\right\}
(f'(0) \varphi'(0) f'(1)\varphi'(1))^{\frac{3}{4}}\times
$$
\begin{equation}
\label{evy}
\times\frac{1}{\sqrt{f'(0)f'(1)}}
\exp\left\{  \frac{\rho^2}{4\sigma ^2}(\frac{f''(0)}{f'(0)}\varphi'(0)- \frac{f''(1)}{f'(1)}\varphi'(1)+
\int \limits_{0}^1 \mathcal{S}ch\,\{f\,,\, \varphi (t)\}\,(\varphi'(t))^{2}\,dt)\right\}
\mu _{\frac{2\sigma}{\rho}}(d\varphi ) )\, d\rho.
\end{equation}

According to Theorem 3, with  $\phi(t)=f(\varphi (t))\,,$ (\ref{evy}) equals to
$$
\frac{1}{\sqrt[4]{f'(0)f'(1)}}
\int \limits _{0}^{+\infty}\
\int \limits _{Diff_+^1([0,1])}\
\delta \left(\frac{1}{\sqrt{f'(1)}}\,\rho\,\sqrt{\phi'(1)}-
\frac{1}{\sqrt{f'(0)}}\,\rho\,\sqrt{\phi'(0)}\right)\times
$$
$$
\times\exp\left\{-\frac{1}{f'(0)\sigma ^2}(a^2+\frac{f''(0)}{4f'(0)} )\,\rho^2\, \phi'(0)-\frac{\sigma ^2}{8\rho^2}\right\}
(\phi'(0) \phi'(1))^{\frac{3}{4}}
\mu _{\frac{2\sigma}{\rho}}(d\phi ) )\, d\rho\,.
$$

From Theorem 2, it follows that
$$
I_{3}\frac{1}{\sqrt[4]{f'(0)f'(1)}}
\int \limits _{C_+ ([0,1])}\
\delta\left(\frac{1}{\sqrt{f'(1)}}u(1)- \frac{1}{\sqrt{f'(0)}}u(0)\right)\,
\exp\left\{-\frac{1}{f'(0)\sigma ^2} (a^2+\frac{f''(0)}{4f'(0)})u^2(0)\right\}
w_{\sigma} (d u)=
$$
$$
=\frac{1}{\sqrt[4]{f'(0)f'(1)}}
\int \limits _{0}^{+\infty}\
\int \limits _{Diff_+^1([0,1])}\
\delta \left(\frac{1}{\sqrt{f'(1)}}\,\rho\,\sqrt{\phi'(1)}-
\frac{1}{\sqrt{f'(0)}}\,\rho\,\sqrt{\phi'(0)}\right)\times
$$
$$
\times\exp\left\{-\frac{1}{f'(0)\sigma ^2}(a^2+\frac{f''(0)}{4f'(0)} )\,\rho^2\, \phi'(0)-\frac{\sigma ^2}{8\rho^2}\right\}
(\phi'(0) \phi'(1))^{\frac{3}{4}}
\mu _{\frac{2\sigma}{\rho}}(d\phi ) )\, d\rho =I_{4}\,,
$$
that proves Lemma 6.

\textbf{Lemma 7.} For any $u\in C([0,\,1])\,,$
$$
\int \limits _{C([0,\,1])}\ \exp\left\{-\imath\int \limits_{0}^{1} u(t)\,y(t)\,,dt  \right\}\,\mu _1(dy)
=\int \limits _{C([0,\,1])}\ \exp\left\{-\imath\int \limits_{0}^{1} u(t)\,y(t)\,,dt  \right\}\,\mu _2(dy)\,.
$$

\textbf{Proof of Lemma 7.}
For $u\in C([0,\,1])\,,$
define $c=\left\|u\right\|$ and $v(t)=c+u(t)\geq 0\,, \ \ \forall t \in [0,1]\,.$
For arbitrary $\alpha\,,\,\beta \in \mathbf{C}\,,\ \
 Re\,\alpha \geq 0\, ,\, Re\,\beta \geq 0\,,$ consider the functions
$$
f_{1}(\alpha, \,\beta)=
\int \limits _{C([0,\,1])}\ \exp\left\{-\int \limits_{0}^1 (\alpha c+\beta v(t)y(t))dt  \right\}\ \mu _{1}(dy)
$$
and
$$
f_{2}(\alpha, \,\beta)=
\int \limits _{C([0,\,1])}\ \exp\left\{-\int \limits_{0}^1 (\alpha c+\beta v(t)y(t))dt  \right\}\ \mu _{2}(dy)\,.
$$
The functions  $f_1(\alpha,\,\beta)\,$ and $f_2(\alpha, \,\beta)$ are continuous on the set
$\bar{\mathbf{C}}_{+}\times \bar{\mathbf{C}}_{+}\,.$ Here, we denote
$$
\mathbf{C}_{+}=\{\alpha \in \mathbf{C}: \,Re\,\alpha > 0 \}\,,\ \bar{\mathbf{C}}_{+}=\{\alpha \in \mathbf{C}: \,Re\,\alpha \geq 0 \}\,.
$$

From Lemma 6, it follows that for $\alpha ,\beta \in \mathbf{R}\,,
 0<\alpha < \frac{1}{8c}\,,\ 0<\beta < \frac{1}{16c}\,,$
$f_{1}(\alpha,\, \beta)=f_{2}(\alpha,\, \beta)\,.$

At $\beta \in (0, \frac{1}{16c})$ fixed, the functions $f_1(\alpha ,\,\beta)$ and $f_2(\alpha ,\,\beta)$ are analytic in $\alpha$ on the set $\mathbf{C}_{+}$, as for any closed smooth curve
$\gamma \subset \mathbf{C}_{+}\,,$ the equalities
$$
\int \limits _{\gamma}f_1(\alpha ,\,\beta)\,d \alpha=
\int \limits _{C([0,\,1])}\ \left(\int \limits _{\gamma}
e^{-\int \limits_{0}^1 (\alpha c+\beta v(t)y(t))dt  }\,d \alpha \right)\,\mu _1(dy)=0,$$
$$\int \limits _{\gamma}f_2(\alpha ,\beta)d \alpha=
\int \limits _{C([0,\,1])}\ \left(\int \limits _{\gamma}
e^{-\int \limits_{0}^1 (\alpha c+\beta v(t)y(t))dt  }\,d \alpha \right)\,\mu _2(dy)=0
$$
are valid.

Thus, for any  $\alpha \in \mathbf{C}_{+}\,,$
$f_{1}(\alpha ,\,\beta) = f_{2}(\alpha ,\,\beta)\,,$ and due to continuity of the functions at $\alpha =-i\,,$
$f_{1}(-i ,\,\beta) = f_{2}(-i ,\,\beta)\,.$

The functions $f_{1}(-i ,\,\beta)$ and  $f_2(-i ,\,\beta)$ are analytic in $\beta$ on the set $\mathbf{C}_{+}\,.$
Therefore, $f_{1}(-i ,\,\beta) = f_{2}(-i ,\,\beta)$  on the set $\mathbf{C}_{+}\,.$
Due to continuity of the functions at $\beta =i\,,\ \ $  $f_{1}(-i ,\,i) = f_{2}(-i ,\,i)\ ,$ that completes the proof of Lemma 7.

Consider the Banach space $E=C([0,\,1])$ with the norm
$\left\| u\right\|=\max \limits _{t \in [0,1]} \left| u(t) \right|$. Denote the dual Banach space as  $E'=(C([0,\,1]))'\,.$

\textbf{Lemma 8.} For any $\varphi \in E'\,,$
$$
\int \limits _{E}\ e^{-i \varphi (y)  }\,\mu _1(dy)
=\int \limits _{E}\ e^{-i \varphi (y)  }\,\mu _2(dy).
$$
\textbf{Proof of Lemma 8.} Consider the function $f \in C(\mathbf{R})$ such that $f(t)=0$  at $\left|t\right|>1$  and
$f(t)=1-\left|t\right|$ at $\left|t\right|\leq 1\,.$

For any positive integers $n\,,\ ( n\geq2 )$ and integers $k$, $0\leq k \leq n\,,$ define the function $f_{n,k} \in E=C([0,\,1])\,,$
for any $t \in [0,1]$ as  $f_{n,0} (t)=\sqrt{2} n\, f(nt)\,,$
$f_{n,n} (t)=\sqrt{2} n\, f(n(t-1))$ and $f_{n,k} (t)= n\, f(nt-k)$ for $1\leq k \leq n-1\,.$

Let $\varphi \in E'$ and  $y \in E=C([0,\,1])\,.$ Define
$$y_n (t)=\sum \limits _{k=0} ^{k=n}f_{n,k} (t) \int \limits_{0}^{1} f_{n,k} (\tau)y(\tau)\,d\tau\, ,\ \ \
u_n (t)=\sum \limits _{k=0} ^{k=n}\varphi(f_{n,k} ) f_{n,k} (t)\,.
$$
For $t \in [\frac{k-1}{n},\frac{k}{n}]\,,$
the equality
$$\left|y_n(t) -y(t)\right|=\int \limits_{0}^{1} f_{n,k-1} (\tau) \left|y(\tau) -y(t)\right|\,d\tau+
\int \limits_{0}^{1} f_{n,k} (\tau) \left|y(\tau) -y(t)\right|\,d\tau
$$
is valid.
Hence,
 $\varphi (y_n )=\int \limits_{0}^1 u_{n} (t)y(t)dt$ and
$\lim \limits_{n \to \infty} \left\|y_n -y\right\|=0\,.$

It means that $\varphi (y)=\lim \limits_{n \to \infty} \varphi (y_n) =
\lim \limits_{n \to \infty} \int \limits_{0}^1 u_{n} (t)y(t)dt$ and
\begin{equation}
\label{Efi1}
\int \limits _{E}\ e^{-i \varphi (y)  }\,\mu _1(dy)=
\lim \limits_{n \to \infty}
\int \limits _{C([0,\,1])}\ e^{-i\int \limits_{0}^1 u_n (t)\,y(t)dt  }\,\mu _1(dy)\,,
\end{equation}
\begin{equation}
\label{Efi2}
\int \limits _{E}\ e^{-i \varphi (y)  }\mu _2(dy)=
\lim \limits_{n \to \infty}
\int \limits _{C([0,\,1])}\ e^{-i\int \limits_{0}^1 u_n (t)\,y(t)dt  }\,\mu _2(dy)\,.
\end{equation}

Due to Lemma 7, (\ref{Efi1}) and (\ref{Efi2}) are equal, that proves Lemma 8.

\textbf{Completion of the proof of Theorem 1.}

Since the Fourier transforms of the measures and, therefore, the measures themselves are equal, and for any continuous bounded function
$F: C_{+}([0,\,1]) \to \mathbf{R}$ we have
$$
\int \limits _{C_+([0,\,1])}\  F(\sqrt{y})\,  \mu _{1}(dy)
=\int \limits _{C_+([0,\,1])}\ F(\sqrt{y})\,\mu _{2}(dy),
$$
Thus,
$$
\int \limits _{C_+ ([0,1])}\ \delta\left(x(1)- x(0)\right)\,x(0)\, F(x)\,
e^{-\frac{a^2}{\sigma ^2}\, x^2(0)}\, w_\sigma (d x)=
$$
$$
=\int \limits _{0}^{+\infty}\ \int \limits _{Diff_{+}^{1}([0,1])}\
\delta\left(\rho\sqrt{\varphi'(1)}- \rho\sqrt{\varphi'(0)}\right)\, \rho\,\sqrt{\varphi'(0)}\, F(x_{\rho,\varphi})\times
$$
$$
\times\exp\left\{-\frac{a^2}{\sigma ^2}\rho^2\varphi'(0)-\frac{\sigma ^2}{8\rho^2}\right\}
(\varphi'(0)\varphi'(1))^{\frac{3}{4}}\,\mu _{\frac{2\sigma}{\rho}}\,(d\varphi )\, d\rho\,.
$$
$$
 x_{\rho,\varphi}(t)=\frac{\rho}{\sqrt{(\varphi^{-1}(t))'}}\,,
\ \ \ \ \ \varphi\in Diff^{1}_{+}([0,\,1])\,,\ \ \ \  0<\rho<+\infty\,.
$$

For $\beta \in \mathbf{R}\,,\ \beta > -1\,,\ g_{\beta}(t)=\frac{(\beta+1)t}{\beta t +1} $ and
$\alpha=a^2 +2\beta ((\beta +1)^{3}-1)\,, $ define
$$
x(t)=g_{\beta}^{-1} u(t)=x(g_{\beta} (t)) \frac{1}{\sqrt{g_{\beta} '(t)}}\,,\
\psi (t)=g_{\beta} (\varphi (t))\,.
$$
In this case, we have
$$
\int \limits _{C_+ ([0,1])}\ \delta\left(\frac{1}{\beta+1}x(1)-(\beta+1) x(0)\right)\,
(\beta+1) x(0)\,F( x)\times
$$
$$
\times\exp\left\{-\frac{\alpha}{\sigma ^2} x^2(0)+\frac{2\beta}{\sigma^2} (-x^2 (0)+\frac1{\beta+1}x^2(1))\right\}\,
w_\sigma (d x)=
$$
$$
=\int \limits _{C_+ ([0,1])}\ \delta\left(u(1)- u(0)\right)\, u(0)\,F(g_{\beta}^{-1} u)\,
\exp\left\{-\frac{\alpha}{\sigma ^2} u^2(0)\right\}\, w_\sigma (d u)=
$$
$$
=\int \limits _{0}^{+\infty}\ \int \limits _{Diff_{+}^{1}([0,1])}\
\delta\left(\rho\sqrt{\psi'(1)}- \rho\sqrt{\psi'(0)}\right)\, \rho\sqrt{\psi'(0)}\,F( g_{\beta}^{-1} x_{\rho,\psi})
$$
$$
\exp\left\{-\frac{\alpha}{\sigma ^2}\rho^2\psi'(0)-\frac{1}{8\rho^2}\right\}
(\psi'(0)\psi'(1))^{\frac{3}{4}}\,\mu _{\frac{2\sigma}{\rho}}\,(d\psi )\, d\rho=
$$
$$
=\int \limits _{0}^{+\infty}\ \int \limits _{Diff_{+}^{1}([0,1])}\
\delta\left(\frac{1}{\beta+1}\,\rho\,\sqrt{\varphi'(1)}-(\beta+1)\,\rho\,\sqrt{\varphi'(0)}\right)\,
(\beta+1)\,\rho\,\sqrt{\varphi'(0)}F( x_{\rho,\varphi})\times
$$
$$
\times\exp\left\{-\frac{\alpha}{\sigma ^2}\,\rho^2\ \varphi'(0)+\frac{\rho^2\beta}{2\sigma^2} (-\varphi'(0)+\frac1{\beta+1}\varphi'(1))-\frac{\sigma ^2}{8\rho^2}\right\}
(\varphi'(0)\varphi'(1))^{\frac{3}{4}}\,\mu _{\frac{2\sigma}{\rho}}(d\varphi ) \,d\rho\,.
$$
Thus,
$$\int \limits _{C_+ ([0,1])}\ \delta\left(\frac{1}{\beta+1}x(1)-(\beta+1) x(0)\right)\,
 x(0)\,F( x)\,
\exp\left\{-\frac{a^2}{\sigma ^2}\, x^2(0)\right\}\,
w_\sigma (d x)=
$$
$$
=\int \limits _{0}^{+\infty}\ \int \limits _{Diff_{+}^{1}([0,1])}\
\delta\left(\frac{1}{\beta+1}\,\rho\,\sqrt{\varphi'(1)}-(\beta+1)\,\rho\,\sqrt{\varphi'(0)}\right)\,
\rho\,\sqrt{\varphi'(0)}\,F( x_{\rho,\varphi})\times
$$
$$
\times\exp\left\{-\frac{a^2}{\sigma ^2}\,\rho^{2}\,\varphi'(0) -\frac{\sigma ^2}{8\rho^2}\right\}\,
(\varphi'(0)\varphi'(1))^{\frac{3}{4}}\,\mu _{\frac{2\sigma}{\rho}}\,(d\varphi )\, d\rho\,.
$$
Denote $\theta =(\beta+1)^{2}\,,$ and rewrite the above equation as
$$
\int \limits _{C_{+} ([0,1])}\ \delta\left(x(1)-\theta x(0)\right)\,
 x(0)\,F( x)\,
\exp\left\{-\frac{a^2}{\sigma ^2}\, x^2(0)\right\}\,
w_{\sigma} (d x)=
$$
$$
=\int \limits _{0}^{+\infty}\ \int \limits _{Diff_{+}^{1}([0,1])}\
\delta\left(\rho\sqrt{\varphi'(1)}-\theta \rho\sqrt{\varphi'(0)}\right)
\rho\sqrt{\varphi'(0)}F( x_{\rho,\varphi})\times
$$
$$
\times\exp\left\{-\frac{a^2}{\sigma ^2}\rho^2\varphi'(0)-\frac{\sigma ^2}{8\rho^2}\right\}
(\varphi'(0)\varphi'(1))^{\frac{3}{4}}\,\mu _{\frac{2\sigma}{\rho}}(d\varphi )\, d\rho\,.
$$
Now we have
$$
\int \limits _{C_+ ([0,1])}\,
 F( x)\,
\exp\left\{-\frac{a^2}{\sigma ^2}\, x^2(0)\right\}\,
w_\sigma (d x)=
$$
$$
=\int \limits _{C_{+} ([0,1])}\ \left(\int \limits _{0}^{+\infty}\ \delta\left(x(1)-\theta x(0)\right)\,
 x(0)\,d \theta \right)\, F( x)\,\exp\left\{-\frac{a^2}{\sigma ^2}\, x^2(0)\right\}\,
w_\sigma (d x)=
$$
$$
=\int \limits _{0}^{+\infty} \ \int \limits _{C_{+} ([0,1])}\ \delta\left(x(1)-\theta x(0)\right)\,
 x(0)\, F( x)\,\exp\left\{-\frac{a^2}{\sigma ^2}\, x^2(0)\right\}\,
w_\sigma (d x)\, d \theta=
$$
$$
=\int \limits _{0}^{+\infty}\, \int \limits _{0}^{+\infty}\ \int \limits _{Diff_{+}^{1}([0,1])}\
\delta\left(\rho\,\sqrt{\varphi'(1)}-\theta\, \rho\,\sqrt{\varphi'(0)}\right)\,
\rho\,\sqrt{\varphi'(0)}\, F( x_{\rho,\varphi})\times
$$
$$
\times\exp\left\{-\frac{a^2}{\sigma ^2}\,\rho^{2}\,\varphi'(0)-\frac{\sigma ^{2}}{8\rho^{2}}\right\}\,
(\varphi'(0)\varphi'(1))^{\frac{3}{4}}\,\mu _{\frac{2\sigma}{\rho}}(d\varphi )\, d\rho\, d \theta=
$$
$$
=\int \limits _{0}^{+\infty}\ \int \limits _{Diff_{+}^{1}([0,1])}\
\left(\int \limits _{0}^{+\infty}\,\delta\left(\rho\,\sqrt{\varphi'(1)}-\theta\, \rho\,\sqrt{\varphi'(0)}\right)\,
\rho\,\sqrt{\varphi'(0)}\, d \theta \right)\,F( x_{\rho,\varphi})\times
$$
$$
\times\exp\left\{-\frac{a^2}{\sigma ^2}\,\rho^2\,\varphi'(0)-\frac{\sigma ^{2}}{8\rho^{2}}\right\}\,
(\varphi'(0)\varphi'(1))^{\frac{3}{4}}\,\mu _{\frac{2\sigma}{\rho}}(d\varphi )\, d\rho=
$$
$$
=\int \limits _{0}^{+\infty}\ \int \limits _{Diff_{+}^{1}([0,1])}\
F( x_{\rho,\varphi})\,
\exp\left\{-\frac{a^2}{\sigma ^2}\,\rho^{2}\,\varphi'(0)-\frac{\sigma ^2}{8\rho^2}\right\}\,
(\varphi'(0)\varphi'(1))^{\frac{3}{4}}\,\mu _{\frac{2\sigma}{\rho}}(d\varphi )\, d\rho\,,
$$
that completes the proof of Main Theorem.

\vspace{0.5cm}

\section{Conclusion: Polar decomposition in two dimensions.}
\label{sec:subst}

\vspace{0.5cm}

In the conclusion, we formulate the theorem on polar decomposition in two-dimensional case.

Define the Wiener measure $w^{\mathbf{C}}_{\sigma}$ on the space of complex-valued continuous functions $C([0, 1], \mathbf{C})$ as the tensor product of the Wiener measures $W_{\sigma}$ for the real and imaginary parts, that is
$$
w^{\mathbf{C}}_{\sigma}(dz)=W_{\sigma}(dx)\ W_{\sigma}(dy)\,,
$$
where $z(t)=x(t)+iy(t)\,,\
z \in C([0, 1],\ \mathbf{C})\,,\ x,y \in C([0, 1])\,.$

Let $C_{\ast}([0, 1], \mathbf{C})=\{ z \in C([0, 1], \mathbf{C})\ :\ z(t)\neq 0 , \, \forall t \in [0, 1] \}\,.$
Then
$$
w^{\mathbf{C}}_{\sigma}(C([0, 1], \mathbf{C}) \setminus C_{\ast}([0, 1], \mathbf{C}))=0\,.
$$

Denote $ S^{1} = \mathbf{R} / \mathbf{Z}\,.$

Now, consider the homeomorphism
$$
L\ :\  \mathbf{R}^{+}\ \times \ Diff^{1}_{+}([0,1])\ \times \ S^{1}\  \times \ C_{0}([0,1])\ \rightarrow C_{\ast}([0,1],\ \mathbf{C})
$$
with
\begin{equation}
   \label{z}
z_{r,\varphi ,\alpha , \eta}=L(r,\varphi ,\alpha , \eta )\,,\ \
z_{r,\varphi ,\alpha , \eta}(t)=\frac{r}{\sqrt{(\varphi ^{-1}(t))'}}
e^{2\pi \alpha i+ i \eta (\varphi ^{-1}(t))}\,,
\end{equation}
where
\begin{equation}
   \label{r,...}
r \in \mathbf{R}^{+}\,,\ \varphi \in Diff^{1}_{+}([0,1])\,,\  \alpha \in S^1\,,\
\eta \in C_{0}([0,1])\,, \ t \in [0,1]\,.
\end{equation}

Then $\forall z \in C_{\ast}([0,1]\,, \mathbf{C})\,,\ (r,\varphi ,\alpha , \eta )=L^{-1} (z)\,,$ where
\begin{equation}
   \label{r,varphi}
r=\left(\int \limits_0^1 \frac{dt}{(\left|z(t)\right|)^2} \right)^{-\frac{1}{2}}\,,\
\varphi ^{-1}(t)=r^2 \int \limits_0^t \frac{d\tau}{(\left|z(\tau)\right|)^2}\,,
\end{equation}
\begin{equation}
   \label{alfa,eta}
 \alpha = \frac{1}{2\pi}Im \,( Ln (z(0)))\,,
\ \ \eta (t) = Im \,( Ln (z(\varphi (t))) - Ln (z(0))) \,.
\end{equation}

Now we define the countably additive measure $\varsigma _\sigma$ on the Banach space $C([0,1],\mathbf{C})$ in the following way:
for any Borel subset $X\subset C([0,1],\,\mathbf{C})$
\begin{equation}
   \label{varsigma}
\varsigma _{\sigma} (X)= \int \limits _{0}^{+\infty}\ \int \limits _{Diff_+^1([0,1])}\
\int \limits _{C ([0,1])}\ \int \limits _{S^1}\
\chi _{X} (z_{r,\varphi ,\alpha , \eta})\
r\ e^{-\frac{\sigma ^2}{4r^2}}\, \varphi '(0) \varphi '(1)  \,
\mu_{\frac{2\sigma}{r}}(d\varphi)\ W^{0}_{\frac{\sigma}{r}}\,(d\eta)\,dr \, d\alpha\,.
\end{equation}
Here, $\chi _{X} $ is the charasteristic function of the set $X\,:\ \chi _X (z)=1$ for $z \in X$ and otherwise
$\chi _X (z)=0\,.$ The variables being related by eqs. (\ref{z}) - (\ref{alfa,eta}).

The following theorem is valid

\textbf{ Theorem 4.} The $w^C_{\sigma}$ and $\varsigma _\sigma$ are equal.

The proof of Theorem 4 will be given in a subsequent paper.

\end{document}